\pdfoutput=1

\documentclass[conference,10pt]{IEEEtran} 

\IEEEoverridecommandlockouts

\usepackage[numbers]{natbib}
\usepackage{amsmath}
\usepackage{nccmath}
\usepackage{multirow}
\usepackage{enumitem}
\usepackage{bigstrut}
\usepackage[tikz]{bclogo}
\usepackage[skins]{tcolorbox}
\usepackage{wrapfig}
\usepackage{booktabs} 
\usepackage{colortbl}
\usepackage{mathtools}
\usepackage{hyperref} 
\usepackage{cleveref}
\usepackage{balance}
\usepackage{multicol}
\usepackage{caption}

\usepackage[flushleft]{threeparttable}
\usepackage[usestackEOL]{stackengine}
\usepackage{xcolor, graphicx}
\usepackage{algorithm}
\usepackage[noend]{algpseudocode}
\usepackage{picture}
\usepackage[final]{listings}
\usepackage[tikz]{bclogo}

\begin{document}

\newcommand\GP[1]{\textcolor{green}{#1}}
\newcommand\TM[1]{\textcolor{blue}{#1}}
\newcommand\AM[1]{\textcolor{blue}{#1}}
\newcommand\MK[1]{\textcolor{red}{#1}}

\lstset{
    language=Python,
    basicstyle=\sffamily\fontsize{2.5mm}{0.8em}\selectfont,
    breaklines=true,
    prebreak=\raisebox{0ex}[0ex][0ex]{\ensuremath{\hookleftarrow}},
    frame=l,
    showtabs=false,
    columns=fullflexible,
    showspaces=false,
    showstringspaces=false,
    keywordstyle=\color{brown}\bfseries\sffamily\fontsize{2.8mm}{0.6em},
    emph={SMOTE, synthetic_samples, minkowski_distance, k, m, r }, emphstyle=\bfseries\color{blue!50!black},
    stringstyle=\color{green!50!black},
    commentstyle=\color{red!50!black}\it,
    numbers=right,
    captionpos=t,
    escapeinside={\%*}{*)}
}

\definecolor{lightgray}{gray}{0.8}

\newenvironment{conditions}
  {\par\vspace{\abovedisplayskip}\noindent\begin{tabular}{>{$}l<{$} @{${}={}$} l}}
  {\end{tabular}\par\vspace{\belowdisplayskip}}

\newcommand{\etal}{et al. }
\newcommand{\IT}[1]{{\bf%
DODGE($\epsilon$)(\ifx*#1$\mathcal{E}$\else#1\fi)}}
\newcommand{\tion}[1]{\S\ref{sect:#1}}
\newcommand{\fig}[1]{Figure~\ref{fig:#1}}

\newcommand{\tbl}[1]{Table~\ref{tbl:#1}}
\newcommand{\bi}{\begin{itemize}[leftmargin=0.4cm]}
	\newcommand{\ei}{\end{itemize}}
\newcommand{\be}{\begin{enumerate}[leftmargin=0.4cm]}
	\newcommand{\ee}{\end{enumerate}}

\setlength\tabcolsep{6pt}



\newenvironment{RQ}{\vspace{2mm}\begin{tcolorbox}[enhanced,width=3.4in,size=fbox,colback=blue!5,drop shadow southeast,sharp corners]}{\end{tcolorbox}}

\newcommand{\quart}[4]{\begin{picture}(70,6)
{\color{black}\put(#3,3){\circle*{4}}\put(#1,3){\line(1,0){#2}}}\end{picture}}
\newcommand{\quartr}[4]{\begin{picture}(70,6)
{\color{black}\put(#3,3){\color{red}\circle*{6}}\put(#1,3){\line(1,0){#2}}}\end{picture}}

\def\tsc#1{\csdef{#1}{\textsc{\lowercase{#1}}\xspace}}
\tsc{WGM}
\tsc{QE}
\tsc{EP}
\tsc{PMS}
\tsc{BEC}
\tsc{DE}




\title{Mining   Workflows  for Anomalous Data Transfers \\[-1.5ex]}
\author{
 \IEEEauthorblockN{Huy Tu\IEEEauthorrefmark{1}, George Papadimitriou\IEEEauthorrefmark{2}, Mariam Kiran\IEEEauthorrefmark{3}, Cong Wang\IEEEauthorrefmark{4}, Anirban Mandal\IEEEauthorrefmark{4}, Ewa Deelman\IEEEauthorrefmark{2}, and Tim Menzies\IEEEauthorrefmark{1}}
 \IEEEauthorblockA{\IEEEauthorrefmark{1}Department of Computer Science, North Carolina State University, Raleigh, USA\\ hqtu@ncsu.edu, timm@ieee.org}
  \IEEEauthorblockA{\IEEEauthorrefmark{2}University of Southern California, Information Sciences Institute, Marina del Rey, CA, USA \\ georgpap@isi.edu, deelman@isi.edu}
   \IEEEauthorblockA{\IEEEauthorrefmark{3}Energy Sciences Network (ESnet), Lawrence Berkeley National Labs, CA, USA\\ mkiran@es.net}
    \IEEEauthorblockA{\IEEEauthorrefmark{4}RENCI, University of North Carolina Chapel Hill, NC, USA\\  cwang@renci.org, anirban@renci.org}
\\[-3.25ex]
}

\markboth{IEEE Mining Software Repositories Conference}%
{Tu \MakeLowercase{\textit{et al.}}: Mining  Scientific  Workflows  for Anomalous Data Transfers for IEEE Journals}

\maketitle
\thispagestyle{plain}
\pagestyle{plain}
\IEEEpeerreviewmaketitle

\begin{abstract}
Modern scientific workflows are data-driven and are often executed on distributed, heterogeneous, high-performance computing infrastructures. Anomalies and failures in the workflow execution cause loss of scientific productivity and inefficient use of the infrastructure. Hence, detecting, diagnosing, and mitigating these anomalies are immensely important for reliable and performant scientific workflows. Since these workflows rely heavily on high-performance network transfers that require strict QoS constraints, accurately detecting anomalous network performance is crucial to ensure reliable and efficient workflow execution. To address this challenge, we have developed X-FLASH, a network anomaly detection tool for faulty TCP workflow transfers. X-FLASH incorporates novel hyperparameter tuning and data mining approaches for improving the performance of the machine learning algorithms to accurately classify the anomalous TCP packets. X-FLASH leverages XGBoost as an ensemble model and couples XGBoost with a sequential optimizer, FLASH, borrowed from search-based Software Engineering to learn the optimal model parameters. X-FLASH found configurations that outperformed the existing approach up to 28\%, 29\%, and 40\% relatively for F-measure, G-score, and recall in less than 30 evaluations. From (1) large improvement and (2) simple tuning, we recommend future research to have additional tuning study  as a new standard, at least in the area of scientific workflow anomaly detection. 






\end{abstract}



\begin{IEEEkeywords}
Scientific Workflow, TCP Signatures, Anomaly Detection, Hyper-Parameter Tuning, Sequential Optimization \\[-2.0ex]
\end{IEEEkeywords}



\section{Introduction}
Computational science today is increasingly data-driven, leading to development of complex, data-intensive applications accessing and analyzing large and distributed datasets emanating from scientific instruments and sensors. Scientific workflows have emerged as a flexible representation to declaratively
express such complex applications with data and control dependencies. Scientific workflow management systems like Pegasus~\citep{deelman-fgcs-2015}, are often used to orchestrate and execute these complex applications on high-performance, distributed computing infrastructure. Examples of these infrastructures include the  Department of Energy Leadership Computing
Facilities; Open Science Grid  \citep{Pordes_2007}; XSEDE \footnote{``Extreme Science \& Engineering Discovery Environment'', xsede.org.}, cloud infrastructures (CloudLab\footnote{``CloudLab'', https://cloudlab.us.}; Exogeni \citep{exogeni}) and national and regional network
transit providers like ESnet\footnote{Lawrence Berkeley National Laboratory, 
ESnet: http://www.es.net.}. 


Orchestrating and managing data movements for scientific workflows within and across this diverse infrastructure landscape is challenging. The problem is exacerbated by different kinds of failures and anomalies that can span all levels of such highly distributed infrastructures (hardware infrastructure, system software, middleware, networks, applications and workflows). Such failures add extra overheads to scientists that forestall or completely obstruct their research endeavors or scientific breakthrough.
At the time of this writing, these problems are particularly acute (the COVID-19 pandemic has stretched the resources used to monitor, maintain and repair the infrastructure).
In particular, scientific workflows rely heavily on high-performance file transfers with strict QoS (Quality of Service: guaranteed bandwidth, no packet loss or data duplication, etc.). Detecting, diagnosing and mitigating for these anomalies is  essential for reliable scientific workflow execution on complex, distributed infrastructures. 

\begin{table*}[!t]
\caption{Hyperparameter tuning options explored by DODGE($\epsilon$) in this paper, drawn from recent SE hyperparameter optimization work~\citep{ghotra2015revisiting,fu2016tuning,agrawal2018better,agrawal2018wrong} then consulting the documentation of a widely-used library
(Scikit-learn by~\cite{pedregosa2011scikit}). Randint,  randuniform and randchoice are  all random functions to choose either integer, float, or a choice among the parameter ranges.}\label{tbl:options}
\vspace{-7pt}

\footnotesize
 
 \begin{tcolorbox}[colback=white]
 \begin{flushleft}
 \vspace{-5pt}
 {\bf DATA PRE-PROCESSORS:} 
\vspace{-10pt}
\begin{multicols}{2}
\bi
\item StandardScaler
\item MinMaxScaler
 \item KernelCenterer
\item Normalizer(norm=a): a = randchoice([`l1', `l2',`max'])
 \item MaxAbsScaler
 \item Binarizer(threshold=a): a =  randuniform(0,100)
\ei
\end{multicols}
\vspace{-13pt}
\bi
 \item RobustScaler(quantile\_range=(a, b)): a, b =  randint(0,50), randint(51,100) 
 \item QuantileTransformer(n\_quantiles=a,  output\_distribution=c, subsample=b). a, b = randint(100, 1000), randint(1000, 1e5);
   c=randchoice([`normal',`uniform']).
    \item SMOTE(a=n\_neighbors, b=n\_synthetics,  c=Minkowski\_exponent). a,b = randit(1,20),randchoice(50,100,200,400). c = randuniform(0.1,5) 
     
\ei
\rule{\linewidth}{0.4pt}

 \vspace{1pt}
\vspace{5pt}
\textbf{LEARNERS:}  

\noindent
\bi
\item DecisionTreeClassifier(criterion=b, splitter=c, min\_samples\_split=a).
    a, b, c= randuniform(0.0,1.0), randchoice([`gini',`entropy']),   randchoice([`best',`random']).
  
\item RandomForestClassifier(n\_estimators=a,criterion=b,  min\_samples\_split=c).  a,b,c = randint(50, 150), randchoice(['gini', 'entropy']),  randuniform(0.0, 1.0) 
     
\item MultinomialNB(alpha=a): a = randuniform(0.0,0.1)
\item KNeighborsClassifier(n\_neighbors=a, weights=b, p=d, metric=c).  a, b,c  = randint(2, 25), randchoice([`uniform', `distance']),  randchoice([`minkowski',`chebyshev']). 
  If c=='minkowski': d= randint(1,15)  else:  d=2

\ei
\vspace{-5pt}

\end{flushleft}
\end{tcolorbox}
\vspace{-17pt}
\end{table*}

\begin{figure}[!t]
\begin{center}
\includegraphics[width=0.49\textwidth, height=2.2in]{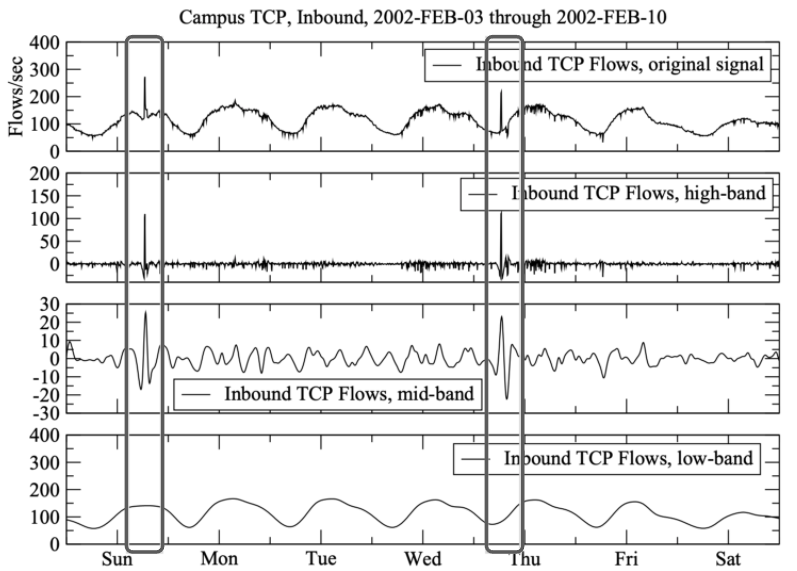}
\vspace{-15pt}
\end{center}
\caption{Example of packet flow signals for a one week period highlighting two anomalies plus high/mid/low decompositions.}
\label{fig:anomalies}
\vspace{-10pt}
\end{figure}

Due to the mission-critical role of such work, this paper seeks ways to build anomaly detectors to specifically explore faulty Transmission Control Protocol (TCP) file transfers similar to those shown in grey boxes of Figure~\ref{fig:anomalies}. According to Papadimitriou et al.~\cite{tcp_indis_19}, such anomalies represent a troubling class of problems. Several research works like~\citep{Lakhina:2004:DNT:1015467.1015492, Mellia:2008:PAT:1410487.1410874} have explored the use of Machine Learning (ML) to detect network anomalies. However, these existing works have mostly employed ``off-the-shelf" ML models, e.g., scikit-learn, without exploring systematic hyperparameter tuning of the ML models themselves. 
Previous research on diverse Software Engineering (SE) problems have shown better learning can be achieved by tuning the control parameters of the ML tools~\citep{dodge, agrawal2018wrong, majumder2018, AGRAWAL2018, fu2016tuning}. Yet, tuning has its own limitations:

\bi

\item \textit{A daunting number of options}: 
Assuming that $R=25$ times\footnote{Why 25? In a 5x5 cross-val experiment, the data set order is randomized five times. Each time, the data is divided into five bins. Then, for each bin, that bin becomes a test set of a model learned from the other bins} we are analyzing $D$ data sets with say $L$ learners and $H$ hyperparameters with each $H$ has continuous or discrete $V$ values to pick, 
then with these settings, hyperparameter optimization with  grid search needs to repeat the experiment millions of times ($R\times L \times D \times H \times V$). For instance, Table \ref{tbl:options} is a sample of the options to explore in this space for $L, H,$ and $V$, which approximately reach billion of choices (assuming each numeric range divides into, say, 10 options).  This is consistent with Agrawal et al.'s report \cite{dodge}.



\item \textit{Slow convergence}:
~\citet{Fu16Grid}'s replication work of~\citet{Tantithamthavorn16} experiment requires 109 days of CPU. A worse case, decades of CPU time were needed by Treude et al. \cite{treude19} to achieve a 12\% improvement over the default settings.

\item \textit{Poorly chosen default configuration}: In ASE 2019's Keynote speech,~\citet{zhou_ase19} remarked that 30\% of errors in cloud environments are due to configuration errors. ~\citet{Jamshidi16} reported for text mining
applications on Apache Storm, the throughput with the worst configuration is 480 times slower than the
throughput achieved by the best configuration.


\ei
Recent work from the SE literature suggests that there exists better state-of-the-art (SOTA) methods to perform  hyperparameter optimization with minimal computational cost~\citep{dodge, agrawal2018wrong, majumder2018, AGRAWAL2018, fu2016tuning}. In research on the DODGE($\epsilon$) algorithm, ~\citet{dodge} reported that with only 30 evaluations by navigating through the output (result) space of the sample of the learners, the preprocessors, and their corresponding parameter choices, DODGE($\epsilon$) outperforms traditional evolutionary approaches. 
In research on the FLASH algorithm,~\citet{flash_vivek} reported that sequential model optimization method can be utilized for software configuration (and possibly hyperparameter tuning).


Drawing inspirations from these work. We designed a network anomaly detection method called X-FLASH, with (1) an ensemble model, XGBoost, and (2) FLASH as a sequential optimizer (to learn the optimal settings for the model). Overall, this paper makes the following contributions:

\bi
\item Investigate the power of hyperparameter tuning to develop anomaly detectors for faulty TCP-based network transfer over SOTA off-the-shelf ML models.
\item First to compare empirically the above two prominent SE-based approaches for hyperparameter tuning.
\item Besides the performance improvement, tuning also changed the conclusion about the most important features for the anomaly detection.

\ei

As a service to other researchers, all  the  scripts and data of this study
are available, on-line\footnote{ \href{https://github.com/msr2021/tuningworkflow}{https://github.com/msr2021/tuningworkflow}}.





\section{Background}

\subsection{Why Study Scientific Workflows?}
Modern computational and data science often involve processing and analyzing vast amounts of data through large scale simulations of underlying science phenomena. With advantages in flexible representation to express complex applications with data and control dependencies, scientific workflows have become an essential component for data-intensive science. They have facilitated breakthroughs in several domains such as astronomy, physics, climate science, earthquake science, biology, among many others~\citep{taylor14_workflows}. 

Reliable and efficient movement of large data sets is essential for achieving high performance in scientific workflow executions. Scientific workflow systems often leverage high-performance networks and networked systems to perform several kinds of data transfers for input data, output data and intermediate data. Hence, the performance and reliability of networks is key to achieving workflow performance. 
As the scientific workflows and the infrastructures supporting them keep increasing both in resource demands and complexity, there is an urgent need for the network to provide high throughput connectivity, in addition to being reliable,  secure, and 99.9\% available. However, there are bound to be anomalies in such large scale systems and applications. Such anomalies are particularly damaging for the scientific research community because (a) poor network performance (e.g., packet loss~\citep{Mellia:2008:PAT:1410487.1410874}) delays scientific discoveries, i.e., negatively impacts scientific productivity, and (b) data integrity issues arising from network errors~\citep{gaikwad16} can jeopardize the validity of scientific results and the reputation of the researchers. Therefore, it is essential to identify and understand these network anomalies early on to allow the network administrator to respond to the anomalies and mitigate the problem.

\vspace{-5pt}
\subsection{Anomaly Detection in Scientific Workflows}

Scientific workflows can take a long time to complete execution because of their scale and complexity comprising a myriad of steps including data acquisition/transformation/pre-processing and model simulation/computing. Therefore, anomalies can be detrimental to both the scientists and the infrastructure providers in terms of lost productivity when long-running workflows fail. Various techniques could be used to predict and detect workflow anomalies. Although domain knowledge could be applied, e.g., ``Execution $E_i$ has failed if it takes longer than $t$ seconds'', this approach is brittle and non-portable between applications and resource types.

Several existing works~\citep{mandal16, gaikwad16, temporal18, herath19, samak11} in end-to-end monitoring of workflow applications and systems are essential building blocks to detect such problems. However, several techniques for anomaly detection are often based on thresholds and simple statistics (e.g., moving averages) ~\cite{jinka2015anomaly}, which fail to understand longitudinal patterns, i.e., relationship between features. Hence, multivariate techniques based on ML are more appropriate to address the anomaly detection problem because they can capture the interactions and relationships between features, as recommended by~\citet{deelman19}.

There is some existing research on the application of ML for scientific workflow anomaly detection. In 2013,~\citet{samak13} employed a Naive Bayes (NB) classifier to predict the failure probability of tasks for scientific workflows on the cloud using task performance data. They found that in some cases, a job destined for failure can potentially be executed successfully on a different resource. Others~\citep{BALA2015980} have compared logistic regression, artificial neural nets (ANN), Random
Forest (RF) and NB for failure prediction of cloud workflow tasks and found that the NB's approach provided the maximum accuracy. In~\citet{buneci08}'s work, the authors have used a k-nearest neighbors classifier to classify workflow tasks into ``Expected'' and ``Unexpected'' categories using feature vectors constructed from temporal signatures of task performance data. Recently,~\citet{herath19} developed RAMP, which is based on using an adaptive uncertainty function to dynamically adjust to avoid repetitive alarms while incorporating user feedback on repeated anomaly detection. In their previous work~\cite{wang-hpec2020}, they have presented a set of lightweight ML-based  techniques, including both supervised and unsupervised algorithms, to identify anomalous workflow behaviors by doing workflow- and task-level analysis. However, none of the above ML-based approaches investigated the possibility of hyperparameter tuning.



\subsection{Model Optimization}

All previous anomaly detection work in scientific workflow lack (1) model optimization and (2) a tuning study. Specifically, this paper is based on the work of~\citet{tcp_indis_19} that applied solely off-the-shelf RF to study faulty TCP file transfers in scientific workflow. It is essential for this study to develop such anomaly detector  with tuning as the backbone. 

Many previous studies  have advised that using data miners without  parameter optimizer is not recommended~\citep{agrawal18, fu2016tuning, spike_jc_19} because: (1) such optimization can dramatically improve performance scores; and (2)  any   conclusions from unoptimized data miner can be changed by new results from the tuned algorithm. For example, Agrawal et al.~\cite{agrawal18} showed how optimizers can improve recall  dramatically by more than 40\%.\hspace{-5pt} Moreover,~\citet{fu2016tuning} showed how optimized data miners  generate different features importance for software defect prediction task. Hence, it is necessary to use data miner using or used by optimizers. However, configuration in the analysis pipeline has numerous problems in their nature that are reported in the \S I.





\begin{table}[!t]
\footnotesize
\begin{center}
\caption{Top 10\% important TCP Statistics features collected by the Tstat tool identified by the state-of-the-art model of \cite{tcp_indis_19} across the data. (S2C: Server to Client, C2S: Client to Server, Both: C2S and S2C)}
\label{tbl:attributes}
\vspace{-3pt}
\begin{tabular} {c|c|l}
Attribute &  Types &  Description \\ \hline
 c\_bytes\_all & C2S	 & Number of bytes transmitted in the \\
 & & payload, including retransmissions \\ 
 c\_pkts\_retx & C2S & Number of retransmitted segments \\ 
 c\_bytes\_retx & C2S  & Number of retransmitted bytes \\ 
 s\_ack\_cnt\_p & S2C	& Number of segments with ACK field \\
 & & set to 1 and no data \\
 durat & ---	& Flow duration since first to last packet \\
 c/s\_first & Both & Client/Server first segment with payload \\
 & & since the first flow segment \\
 c/s\_last  & Both & Client/Server last segment with payload \\
 & & since the first flow segment \\
 c/s\_first\_ack & Both & Client/Server first ACK segment (without  \\
  & & SYN) since the first flow segment	 \\
 c/s\_rtt\_avg & Both & Average RTT computed measuring the time \\
 & & elapsed between the data segment and \\
 & & the corresponding ACK\\
c/s\_rtt\_min & Both & Minimum RTT observed in connection lifetime \\
c/s\_rtt\_max & Both & Maximum RTT observed in connection lifetime \\
 \hline
 
\end{tabular}
\end{center}
\vspace{-15pt}
\end{table}

Solving these configurations is not limited to software systems and hyperparameter optimization in ML but also for cloud computing and software security. In cloud computing, different analytic jobs have diverse behaviors and resource requirements, choosing the correct combination of  virtual machine type and cloud environment can be critical  to optimize the performance of a system while
minimizing cost~\citep{Alipourfard17_cherry,venka16_ernest, Zhu17_bestconfig, Dalibard17_boat, Yadwadkar14}. In security of cloud computing, such problems of how to maximize conversions on landing pages
or click-through rates on search-engine result pages~\citep{wang18_web, zhu17_click, Hill17_bandit} has gathered interest.

\subsection{Faulty TCP Case Study}\label{subsec:tcp_study}

\begin{wrapfigure}{r}{0.2\textwidth}
\vspace{-10pt}

\begin{center}
\hspace{-15pt}\includegraphics[width=0.22\textwidth]{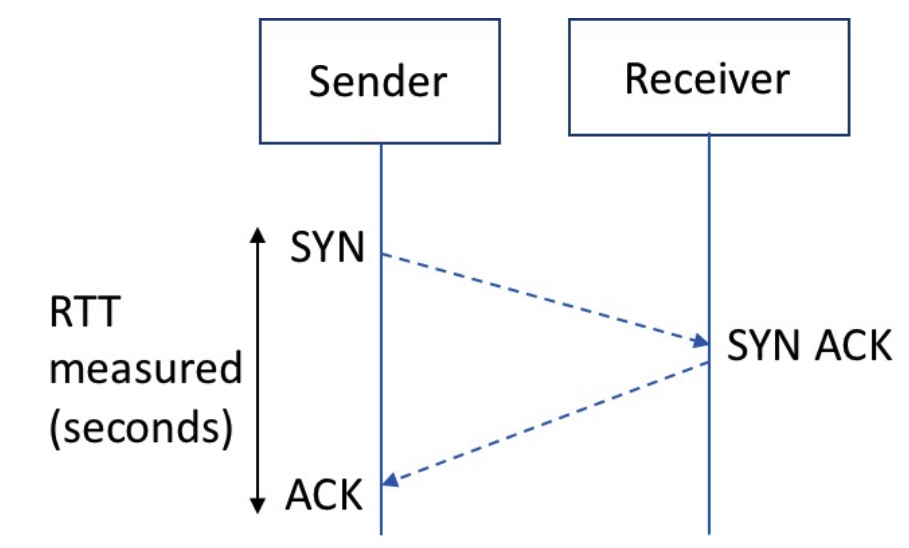}
\end{center}
\caption{Round Trip Time measured in TCP's three way
communication.}
\label{fig:TCP}
\vspace{-10pt}
\end{wrapfigure}


In this work, we analyze anomalous network transfers by
utilizing data collected using TCP statistics (Tstat) \citep{tstat}, which is a tool to
collect TCP traces for transfers. How TCP
works can be demonstrated in Figure \ref{fig:TCP} as follows: the server sends a packet to a client (with the load in bytes), when the client receives the packet, it sends an
``acknowledgment'' (ACK) signal back. The round trip time (RTT) is the total time from sending the package to receiving 
the ACK. Time windows is a measure that is used by the TCP
protocol to allow servers to wait for
the ACK, before deciding to resend the packet again. 



Other than package loss, there is significant effort to recognize
 overflowing buffers~\citep{host, lte} and
 commonly occurring network anomalies
 seriously impact user experience and also affecting the clients' work negatively as mentioned in \S II.A. Described by \cite{dapper_17, survey_anom16, passive_08}, three common network anomalies targeted by this study are: 

\bi
\item Packet Loss happens when one or more packets fail
to reach their destination. These could either be caused
by errors in transmission or too much congestion on link,
causing routers to randomly drop packets.
\item Packet Duplication happens when the sender re-
transmits packets, thinking that the previous packets have
not reached their destination. This can be commonly observed 
when packet losses happen and retransmits increase.
\item Packet Reordering happens when arrival order
of packets or sequence number is completely out-of-
order. In the case of real-time media
streaming application, it is particularly relevant to show network instability.
\ei

Collectively, Tstat traces contain 133 variables per packet on both server and client sides so TCP protocol can ensure the packages are delivered reliably. These features are listed in details on Tstat’s documentation \footnote{http://tstat.polito.it/}. Table \ref{tbl:attributes} reports the top 10\% features ranked by their importance shared across the data by applying the state-of-the-art work by~\citet{tcp_indis_19}. It is mission-critical to understand what attributes are essential to the model to make decisions on classifying the right type of anomalies. Yet, the tuning study here showed that the important attributes reported  before tuning and after tuning change significantly. This indicates that previous anomalies detection work reported misleading key features to the system managers or scientists which can cost them extra resource to debug. Accordingly, our proposed solution, X-FLASH  does inform the right key features to the system experts to take appropriate action and prevent future network anomalies.



TCP provides reliable and error-checked delivery of a data
stream between senders and receivers. Research efforts have been focusing on TCP extensions as variants to allow improvement
of various network anomalies and enable congestion control. Jacobson et al.~\cite{jacobson88} established implementations of the modern TCP. Since that seminal work, some TCP variants are introduced to prioritize throughput over
loss prevention. In this paper scope, we specifically focus on four of these variants: Cubic, Reno, Hamilton, and BBR. For more information regarding these variants, please see~\cite{tcp_indis_19}.

\section{Software Configuration Optimization}

The case was made above that (a)~anomalies detection in scientific workflow is mission-critical, (b)~previous studies lack optimization for their analytics pipelinem, and (c)~tuning is a daunting task that requires careful attention per domain. Therefore, X-FLASH is designed to include tuning in the data mining pipeline for anomalies detection in scientific workflow. 
This section described how FLASH and DODGE($\epsilon$) can tune the data mining pipeline for a better scientific workflow anomalies detection.  

\subsection{Core Problem}

The problem can be described starting with a configurable data miner has a set $X$ of configurations $x \in X$. Let  $x_i$ represent the $ith$ configuration of a data miner method. $x_{i,j}$ represent the $jth$ configuration option
of the configuration $x_i$. In general, $x_{i,j}$ indicates either an
(i) integer variable or a (ii) Boolean variable. The configuration space (X) represents all the valid configurations of a data miner tool. The configurations are also referred to as independent variables ($x_i$) where $1 \leq i \leq |X|$, has one (single-objective) corresponding performance measures $y_{i,k} \in Y$ indicating the $1 \leq kth \leq m$ objective (dependent variable). In our setting, the cost of optimization technique is the total number of iteration required to find the best configuration settings. 

We consider the problem of finding a good configuration, $x^*$, such that $f(x^*)$ is less than other configurations in X. Our objective is to find $x^*$ while minimizing the number of iterations and measurements. 

\subsection{Overview}

The heart of this problem is to optimize the analytical results (performance) with the knowledge at hand while minimizing the iterations (time) for the model to converge. In SE literature, the solution can be seen with evolutionary optimization (based on mutating existing configurations). However, according to Nair et al.~\cite{flash_vivek} and Agrawal et al.~\cite{agrawal2018better, agrawal2018wrong, dodge}, such optimization can be cost inefficient, slow convergence, and poor performance. Hence, research in software configuration in
the last decade has explored non-EA methods including Sequential Model-Based optimization, and $\epsilon$-dominance.

\begin{figure*}[!t]
\includegraphics[width=1\textwidth, height=1in]{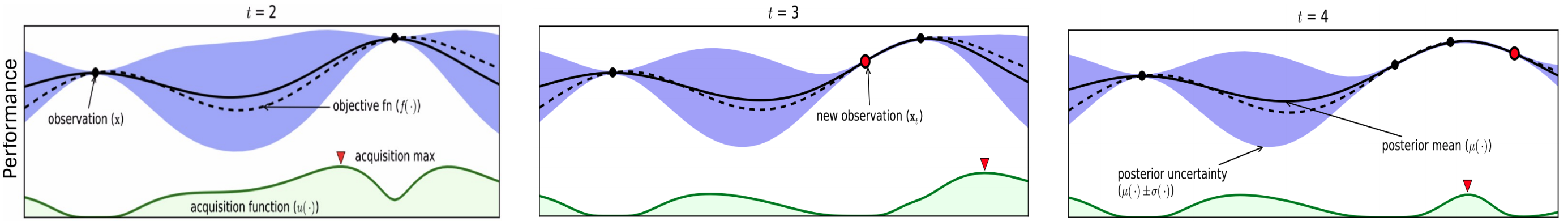}
\caption{An example of Sequential Model-based method’s working process from \citet{brochu10}. The figures show a Gaussian
process model (GPM) of an objective function over four iterations of sampled values. Green shaded plots represent
acquisition function. 
The value of the acquisition function is high where the GPM predicts larger objective and where
the prediction uncertainty (confidence) is high such points (configurations in our case) is sampled first. Note that the
area on the far left is never sampled even when it has high uncertainty (low confidence) associated.}
\label{fig:smbo}
\vspace{-15pt}
\end{figure*}

For those reasons, research in this area in
the last decade has explored non-EA methods for software
configuration.  

\subsubsection{Sequential Model-Based optimization}

FLASH is a variant of Sequential Model-based Optimization (SMBO) whose core concept is ``given what we know about the problem, what should we do next?''.  To illustrate this, consider Figure \ref{fig:smbo} of SMBO. The bold black line represents the actual performance function ($f$, which is unknown in our setting) and the dotted
black line represents the estimated objective function (in the
language of SMBO, this is the \textit{prior}).  The optimization starts with two points (t=2).
At each iteration, the acquisition function is maximized
to determine where to sample next.
A model is built on the points and these evaluated measurements as the prior belief. This model can then learn where to sample next and find extremes of an unknown objectives. A posterior is then defined and captured as our updated belief in the objective function or surrogate model. The purple regions
represent the configuration or uncertainty of estimation in a
region—the thicker that region, the higher the uncertainty. The green line in that figure represents the acquisition
function. The acquisition function is a user-defined strategy, which takes into account
the estimated performance measures (mean and variance)
associated with each configuration. The chosen sample (or
configuration) maximizes the acquisition function ($argmax$).
This process terminates when a predefined stopping condition is reached which is related to the budget associated
with the optimization process.

Gaussian process models are often the surrogate model of choice in the literature. Yet, building GPM can be very challenging since (1) GPM can be very fragile to the parameters setting and (2) GPM do not scale to high dimensional data as well as large data set (i.e., large option space).  Therefore,~\citet{flash_vivek} proposed the SMBO's improvement, i.e., FLASH: 

\bi

\item FLASH models each objective as a separate Classification and Regression Tree (CART) model. Nair et al. reported that the CART
algorithm can scale much better than other model
constructors (e.g., Gaussian Process Models). 

\item FLASH replaces the actual evaluation of all combinations of parameters(which can be very slow) with a \textit{surrogate evaluation}, where the CART decision trees are used to guess the objective scores (which is very fast). Such guesses may be inaccurate but, as shown by~\citet{flash_vivek}, such guesses can rank guesses in (approximately) the same order as that generated by other, much slower, methods~\citep{nair2017using}.

\ei


FLASH can be executed as follow:
\begin{enumerate}[start=1,label={\bfseries Step \arabic*}, leftmargin=1.1cm]

\item \textbf{Initial Sampling:} A sample of predefined configurations from the option space is evaluated. The evaluated configurations are removed from the unevaluated pool.

\item \textbf{Surrogate Modeling:} The evaluated configurations and the corresponding performance measures are then used to build CART models.

\item \textbf{Acquisition Modeling}: The acquisition
function accepts the generated surrogate model (or models)
and the pool of unevaluated configurations (uneval configs)
to choose the next configuration to measure. 

For multi-objective problems, for each configuration $x_i$, $N$ (random projections) vectors $V$ of length $o$ (objectives) are generated with:
\begin{itemize}
    \item Guess it's performance score $y_{i,j}$ using CART.
    \item Compute its mean weight as: 
    \begin{center}
    $mean_i = \frac{1}{N} \sum_n^N \sum_j^o(V_{n,j} \cdot y_{i,j})$   
    \end{center}
    
    \item if $mean_i > max_{current}$ then $best = x_i$. 
\end{itemize}

\item \textbf{Evaluating}: The configuration chosen by the acquisition function is evaluated and removed from the configuration candidates pool.

\item \textbf{Terminating}: The method terminates once it runs out of a  predefined budget.
\end{enumerate}



FLASH was invented for the software configuration problem as it performed faster than more traditional optimizers such as Differential Evolution~\citep{Storn1997DifferentialE} or NSGA-II~\citep{996017}. FLASH is an improvement of SMBO and was chosen as the optimizer component for X-FLASH in a few technical areas including:

\bi 

\item For surrogate modeling, CART is replaced with GPM. GPM would have taken $O(M^3)$ time whereas CART is a bifurcating algorithm that would only take $O(M\cdot N^2)$ where M is the size of the training dataset and N
is the number of attributes. 

\item FLASH's acquisition function uses Maximum Mean. By assuming that the greatest mean might contain the
values that most extend to the desired maximal (or
minimal) goals. It cut the runtime from $O(M^2)$ to only $O(M)$.

\item GPM assumed smoothness where configurations that are close to each other have similar performance. However, CART makes no assumption that neighboring
regions have the same properties.

\ei

\subsubsection{DODGE and $\epsilon$-Dominance}

\begin{figure}[!b]
\vspace{-20pt}
\centering
\includegraphics[width=0.6\linewidth]{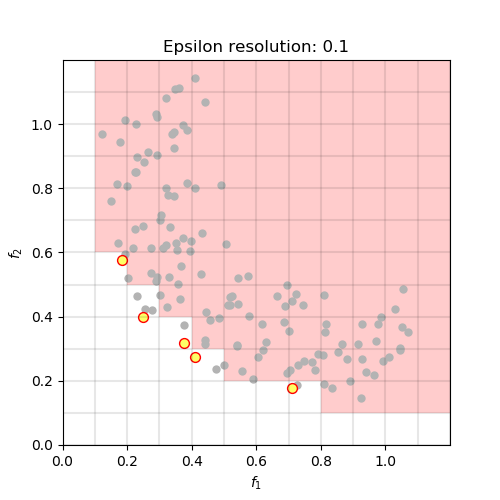}
\caption{An example of Pareto frontier \cite{shu2019improved}. Objectives $f_1$ and $f_2$ are both to be minimized. $\epsilon$ value is set to 0.1. The gray dots represent the feasible solutions. The yellow dots represent the Pareto optimal solutions. }
\label{fig:pareto}
\end{figure}

In 2005,~\citet{deb2005evaluating} proposed partitioning the output space of an optimizer into $\epsilon$-sized grids. In multi-objective optimization (MOO), a solution is said to \textit{dominate} the other solutions if and only if it is better in at least one objective, and no worse in other objectives. A set of optimal solutions that are not dominated by any other feasible solutions form the \textit{Pareto frontier}. Figure~\ref{fig:pareto} is an example of the output space based on $\epsilon$-dominance. The yellow dots in the figure form the Pareto frontier.

Deb's principle of $\epsilon$-dominance is that if there exists some $\epsilon$ value below which is useless or impossible to distinguish results, then it is superfluous to explore anything less than $\epsilon$~\citep{deb2005evaluating}. Specifically, consider distinguishing the type of anomalies discussed in this paper, if the performances of two learners (or a learner with various parameters) differ in less than some $\epsilon$ value, then we cannot statistically distinguish them. For the learners which do not significantly improve the performance, we can further reduce the attention on them.

~\citet{dodge} successfully applied $\epsilon$-dominance to some SE tasks such as software defect prediction and SE text mining. Their proposed approach, named DODGE($\epsilon$), was a tabu search, i.e., if some settings arrive within $\epsilon$ of any older result, then DODGE($\epsilon$) marked that option as ``to be avoided''.

 A tool for software analytics, DODGE($\epsilon$)   needed just a few dozen evaluations to explore billions of    configuration options 
 for (a)~choice of learner, for (b)~choice of pre-processor, 
 and for (c)~control parameters for the learner and pre-processor. These configurations, when combined together, make up billions of options that are reported in Table 1 of \cite{dodge}.
 DODGE($\epsilon$) executed by:
 \vspace{5pt}
 \be
 \item
 Assign     weights $w=0$ to   configuration options.
 \item
 $N=30$ times repeat:
 \be
 \item
 Randomly pick  
  options, favoring   those with most weight;
  \item
  Configuring and executing data pre-processors and learners using those options;
  \item
  Dividing output scores into regions of size $\epsilon=0.2$;
 \item
 if some new configuration has scores with  $\epsilon$ of  
  prior configurations then...
  \bi
  \item
  ...reduce the weight of those configuration options $w=w-1$; 
  \item Else, add to their weight with $w=w+1$.
  \ei
  \ee
  \item Return the best option found in the above.
  \ee
  Note that after  Step 2d, the choices made in subsequent Step 2a will avoid options that result in $\epsilon$ of 
 other observed scores.
 
 Experiments with DODGE($\epsilon$) found that best learner performance plateau after just $N=30$ repeats of Steps 2-5.
 To explain this result,~\cite{dodge} note that for a range of software analytics tasks,  the outputs of a learner divide into  only a handful of equivalent regions.
For example, when a software analytics task
  is repeated 10 times, each time with 90\% of the
data, then the  observed performance scores   (e.g., recall, false alarm)
can vary by 5 percent, or more. Assuming normality, 
then  scores less than $\epsilon=1.96*2*0.05=0.196$ are statistically indistinguishable. Hence, for learners    evaluated on (say) $N=2$  scores, 
  those  scores   effectively divide into just \mbox{ $C=\left(\frac{1}{\epsilon = 0.196}\right)^{N=2}=26$} different regions.  Hence, it is hardly surprising that a few dozen repeats of Steps 2-5 were enough to explore a seemingly very large space of options.







\section{Methodologies}

 While the above DODGE($\epsilon$) and FLASH algorithms
 have been shown to work well for analytics tasks in
 software engineering (e.g., effort estimation, bug location etc). 
 These algorithms have not been successfully deployed
 outside the realm of SE.
 Accordingly,
 the rest of this paper tests if DODGE($\epsilon$) and/or FLASH 
 work well for anomaly detection for faulty Transmission Control Protocol.

\subsection{Data}
The data here is adopted from the SOTA TCP anomaly detection~\cite{tcp_indis_19}. It includes two sets of datasets of Mice\&Elephant Flows and 1000 Genome Workflow where each set include four datasets corresponding to four TCP variants (Hamilton, BBR, Reno, and Cubic) under \textit{normal} or \textit{anomalous} conditions (loss, duplicate, and reordering). A summary of both sets is captured in Table~\ref{tbl:data} which depicts the number of collected flows across anomaly types and TCP variants.

\subsubsection{Mice and Elephant Flows} ExoGENI
testbed~\cite{exogeni} is used to generate this labeled set of data. ExoGENI is a federated cloud testbed designed for experimentation and computational tasks. It is orchestrated over a set of independent cloud sites located across US and connected via national research circuit providers through
their programmable exchange points. Mice flows were aimed for 1000 SFTP
transfers with a transfer size between 80 MB and 120 MB, the
link bandwidth is set to 1 Gbps among all the nodes.
Elephant flows were aimed for 300
SFTP transfers with a transfer size between 1 and 1.2 GB, the
link bandwidth is set to 100 Mbps among all the nodes.

\begin{table}[!t]
\scriptsize
\begin{center}
\caption{Number of flows generated at Data Node across TCP variants (H-Hamilton, B-BBR, R-Reno, \&  C-Cubic) under
\textit{normal} or \textit{anomalous} settings (loss, duplicate, \& reordering)  for  Mice\&Elephant Flows and 1000 Genome Workflow.}
\vspace{-1pt}
\label{tbl:data}
\begin{tabular} {p{.11\linewidth}|c|c|c|c|c|c|c|c}
&  \multicolumn{4}{ c| }{Mice\&Elephant Flows} & \multicolumn{4}{ c }{1000 Genome Workflow} \\ \hline
 Type & H & B & R & C & H & B & R & C \\ \hline
 Normal  & 1304 & 1304 & 1304 & 1304 & 550 & 508 & 532 & 528    \\ \hline
 Loss & 3994 & 3975 & 3989 & 3995  & 6110 & 10588 & 1212 & 1721   \\ \hline
 Duplicate & 2616 & 2615 & 2616 & 3778 & 1111 & 1016 & 1097 & 1083  \\ \hline
  Reordering & 3830 & 2612 & 2612 & 2614 & 1141 & 1019 & 1078 &  1067   \\ 
\end{tabular}
\end{center}
\vspace{-15pt}
\end{table}

\subsubsection{1000 Genome Workflow Transfers}

\begin{figure}[!b]
\vspace{-5pt}
	\centering
	\includegraphics[width=\linewidth]{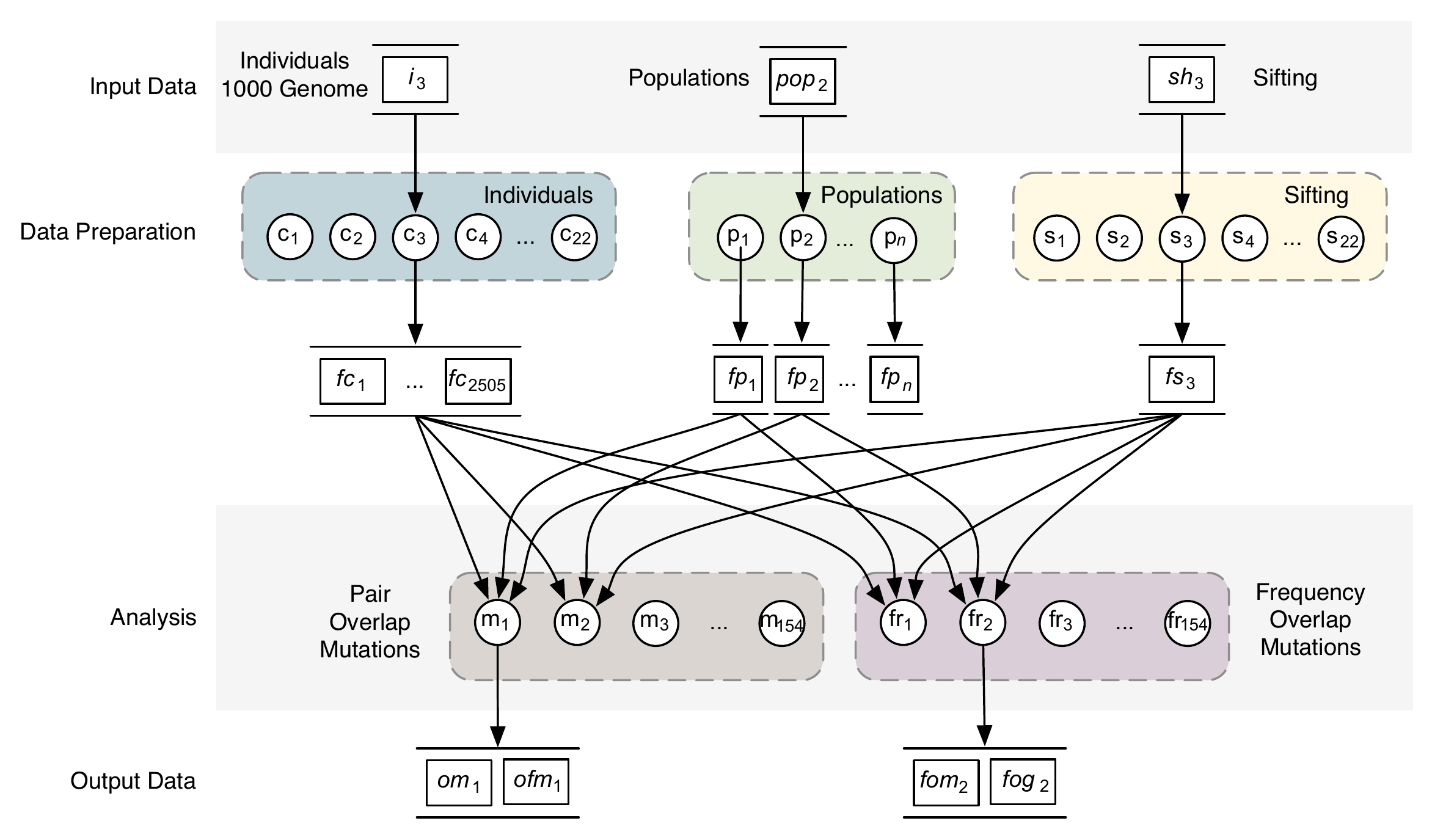}
	\caption{The overview of 1000 Genome sequencing analysis workflow.}
	
	\label{fig:1000genome}
\end{figure}

The data used in this study are comprised of network traces produced by the 1000 Genome Pegasus workflow. This science workflow is inspired by the 1000 genomes project which provides a reference for human variation, having reconstructed the genomes of 2,504 individuals across 26 different populations~\citep{auton15}. The version of the 1000 Genome workflow used (Figure~\ref{fig:1000genome}) is composed of five different tasks: (1)~\textit{individuals} – fetches and parses the Phase 3 data from the 1000 genomes project per chromosome; (2)~\textit{populations} -- fetches and parses five super populations (African, Mixed American, East Asian, European, and South Asian) and a set of all individuals; (3)~\textit{sifting} -- computes the SIFT scores of all of the SNPs (single nucleotide polymorphisms) variants, as computed by the Variant Effect Predictor; (4)~\textit{pair overlap mutations} -- measures the overlap in mutations (SNPs) among pairs of individuals; and (5)~\textit{frequency overlap mutations} -- calculates the frequency of overlapping mutations across subsamples of certain individuals.
During the lifetime of the workflow, most of the data needs to be staged in during the first part of the workflow (\textit{individuals}) with many smaller transfers following to accommodate the execution of the rest of the processing tasks. In their experiments, they included a comprehensive list of TCP  and conducted workflow transfers under normal and various anomalous conditions in~\ref{subsec:tcp_study}. 




\subsection{Data Miners}

Hyperparameter optimizers (i.e., FLASH) tune the settings of
data miners. This section describes such data miner candidates to be tuned in this study.

\subsubsection{CART and RF}

We use CART to recursively build decision trees to
find the features that reduce most of {\em entropy}, where a higher entropy indicates less ability to draw conclusion from the data being processed~\citep{CART}.
Using CART as a sub-routine, our
Random Forest method builds many trees,
each time with different subsets of the data rows $R$ and columns $C$\footnote{Specifically, using $\log_2{C}$ of the columns, selected at random.}. 
Test data is then passed across all $N$ trees and the conclusions are determined (say) a majority vote across all the trees~\citep{Breiman2001}. Holistically, RF is based on bagging (bootstrap aggregation) which averages the results over many decision trees from sub-samples (reducing variance). Both are popular in the field of ML and implemented in popular open-source toolkit Scikit-learn by \cite{pedregosa2011scikit}.

We choose to use CART for its explainability and efficiency, as discussed in \cite{flash_vivek, xia2019sequential, spike_jc_19}. RF was used in the previous study for this same problem for its advantages of performance~\citep{kiran_jmlr}. However, we will show later in this paper that RF without optimization is not enough. 

\subsubsection{XGBoost}

\textit{Gradient Boosting} is chosen as a model for it's advantages of reducing both variance and bias. It is an ensemble model which involves:

\bi
\item Boosting builds models from individual so called ``weak learners'' in an iterative way. The individual models here are not built on completely random subsets of data and features but sequentially by putting more weight on instances with wrong predictions and high errors (reducing biases). 
\item The gradient is a partial derivative of our loss function - so it describes the steepness of our error function in order to minimize error in the next iteration.  
\ei

Gradient Boosting reduces the variances with multiple models (similar to bagging in RF) and also reduces bias with subsequently learning from previous step (boosting). XGBoost is an improved Gradient Boosting method by (1) computing second-order gradients, i.e., second partial derivatives of the loss function (instead of using CART as the loss function); and (2) advanced regularization (L1 \& L2)~\citep{xgboost}.

\subsection{Evaluation Metrics}

The problem studied in this paper is a multiclass classification task with four classes (1 normal class and 3 anomalous classes). The performance of such multiclass classifier can be assessed via a confusion matrix as shown in Table \ref{tbl:cmatrix} where each class is denoted as $C_i$. 

Further, ``false'' means the learner got it wrong and ``true'' means the learner correctly identified
a positive or negative class. The four counts include True Positives (TP), False Positive (FP), False Negative (FN) and True Negative (TN). 

\begin{table}[t!]
\small
\begin{center}
\caption{Results Matrix of Multiclass Classification}
\vspace{-2pt}
\label{tbl:cmatrix}
\begin{tabular} {@{}cc|c|c|c|c|l@{}}
& & \multicolumn{4}{ c| }{Actual} \\ \cline{2-6}
& \multicolumn{1}{ c| }{Prediction} & C1 & C2 & C3 & C4  \\ \cline{2-6}
& \multicolumn{1}{ c| }{C1} & $\mathit{TP_{11}}$ & $\mathit{FN_{12}}$ & $\mathit{FN_{13}}$ & $\mathit{FN_{14}}$ & \\ \cline{2-6}
& \multicolumn{1}{ c| }{C2} & $\mathit{FP_{21}}$ & $\mathit{TP_{22}}$ & $\mathit{FN_{23}}$ & $\mathit{FN_{24}}$ & \\ \cline{2-6}
& \multicolumn{1}{ c| }{C3} & $\mathit{FP_{31}}$ & $\mathit{FP_{32}}$ & $\mathit{TP_{33}}$ & $\mathit{FN_{34}}$ & \\ \cline{2-6}
 & \multicolumn{1}{ c| }{C4} & $\mathit{FP_{41}}$ & $\mathit{FP_{42}}$ & $\mathit{FP_{43}}$ & $\mathit{TP_{44}}$ &  \\ 
\end{tabular}
\end{center}
\vspace{-20pt}
\end{table}

Due to the multiclass and anomalies detection nature with no imbalanced class issue observed, we want to make sure all classes are treated fairly. A macro-average is preferred to compute each metric independently for each class $C_i$ and then take the average. Ling et al.~\cite{1ling_auc} and Menzies et al.~\cite{precision_menzies} had warned us against accuracy and precision as evaluation metrics even when the original work employed accuracy.   
Therefore, we used 3 macro-average measures, i.e., recall, F-measure (a harmonic mean of precision and recall), and G-score (a harmonic mean of recall and false-alarm rate, or FAR) to evaluate the learners that are calculated as below:
\bi 
\item \footnotesize $Precision = \frac{\sum_{i=1}^{L}\frac{TP_{i}}{TP_{i} + FP_{i}}}{L}$

\item \footnotesize $FAR = \frac{\sum_{i=1}^{L}\frac{FP_{i}}{TP_{i} + TN_{i}}}{L}$

\ei

\begin{ceqn}

\begin{equation}
\scriptsize
Recall = \frac{\sum_{i=1}^{L}\frac{TP_{i}}{TP_{i} + FN_{i}}}{L}
\end{equation} 

\begin{equation}
\scriptsize
F-measure = \frac{2 * Precision * Recall}{Precision + Recall}
\end{equation}

\begin{equation}
\scriptsize
\mathit{G-score} = \frac{2 \cdot \mathit{Recall} \cdot \mathit{(1 - FAR)} }{\mathit{Recall} + (1 - \mathit{FAR})}
\end{equation}

\end{ceqn}



\subsection{Statistical Testing}


We compared our results using Scott-Knott method, which
sorts results from different treatments, and then splits them to maximize the expected value of differences in the observed performances
before and after divisions. For lists $l,m,n$ of size $\mathit{ls},\mathit{ms},\mathit{ns}$ where $l=m\cup n$, the ``best'' division maximizes $E(\Delta)$; i.e., the delta in the expected mean value before and after the split: 

\[E(\Delta)=\frac{ms}{ls}abs(m.\mu - l.\mu)^2 + \frac{ns}{ls}abs(n.\mu - l.\mu)^2\]

Scott-Knott then checks if that ``best'' division is actually useful. To implement that check, Scott-Knott would apply some statistical hypothesis test $H$ to check if $m, n$ are significantly different (and if so, Scott-Knott then recurses on each half of the ``best'' division). For this study, our hypothesis test $H$ was a conjunction of statistical significance test and an effect size test. Specifically, significance test here  is non-parametric bootstrap sampling, which is useful for detecting if two populations differ merely by random noise, cliff's delta \citep{mittas2013ranking,ghotra2015revisiting}. Cliff's delta quantifies the number of difference between two lists of observations beyond p-values interpolation~\citep{macbeth2011cliff}. The division passes the hypothesis test if it is not a ``small'' effect ($Delta \geq 0.147$). The cliff's delta non-parametric effect size test explores two lists $A$ and $B$ with size $|A|$ and $|B|$:

\begin{equation}
    Delta = \frac{\sum\limits_{x \in A} \sum\limits_{y \in B} \left\{ \begin{array}{l}
                    +1, \mbox{   if $x > y$}\\
                    -1, \mbox{   if $x < y$}\\
                    0,  \mbox{   if $x = y$}
                \end{array} \right.}{|A||B|}
\end{equation}

In this expression, cliff's delta estimates the probability that a value in list $A$ is greater than a value in list $B$, minus the reverse probability~\citep{macbeth2011cliff}. This hypothesis test and its effect size is supported by Hess and Kromery~\citep{hess2004robust}.


\begin{figure}[!t]
\vspace{-20pt}
\centering
\includegraphics[width=\linewidth, height=2in]{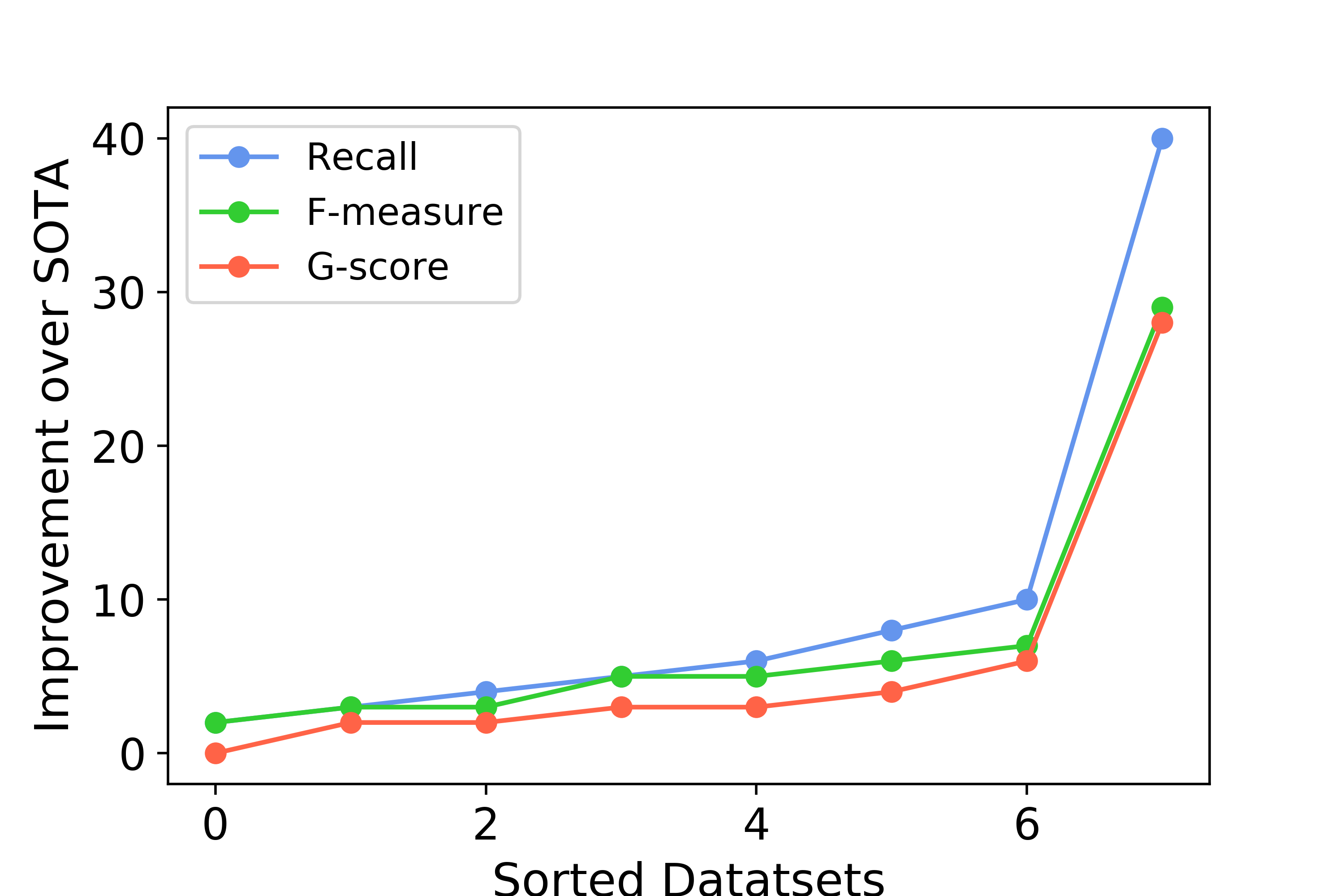}
\caption{Sorted relative median delta for each performance across all datasets between X-FLASH and RF (SOTA work's choice of data miner method).} 
\label{fig:delta}
\vspace{-15pt}

\end{figure}

\subsection{Test Rig}
 
 We applied $k$-fold cross-validation, with $k = 10$ to randomly partition the data into k equal sized subsamples. A single subsample among them is retained for testing, and the remaining $k - 1$ subsamples are used for tuning and validation with the proportions of 80\% and 20\% respectively. FLASH and DODGE($\epsilon$) are applied on tuning dataset and validated on validation dataset before evaluated on the test dataset. The cross-validation process is then repeated $k$ times. The advantage of this method over repeated random sub-sampling is that all observations are used for both training and testing. 

\begin{table}[!b]
\footnotesize
\begin{center}
\caption{Default and median tuned configurations of four XGBOOST's parameters when tuning for a specific metric (recall, F-measure, and G-score) across eight datasets.}
\vspace{-5pt}
\label{tbl:params}
\begin{tabular} {p{.16\linewidth}|c|c|c|c}

 Parameter & Defaut & Recall & F-measure & G-score \\ \hline
 max\_depth  & 3 & 12 & 17 & 19    \\ 
 learning\_rate & 0.1 & 0.53 & 0.53 & 0.58   \\ 
 \#\_estimators & 100 & 107 & 105 & 106  \\ 
 booster & gbtree & dart & gbtree & dart \\
\end{tabular}
\end{center}
\vspace{-15pt}
\end{table}

\section{Results}



\begin{table*}[!htp]
\scriptsize
\caption{Recall, F-measure, and G-score (higher the better) results are reported between all the methods that are ranked by Scott-Knott statistical tests. Medians and IQRs (delta between 75th percentile and 25th percentile, lower the better) are calculated for easy comparisons. The best performing one is noted as a gray cell for each metric per column. The column \textbf{\#Best} shows the number of projects each treatment performs the best in.}
\centering
\setlength\tabcolsep{4pt}
\begin{tabular}{p{.065\linewidth}|p{.1\linewidth}|c|c|c|c|c|c|c|c|c|c|c}
\multirow{2}{*}{\textbf{Metrics}}   & \multirow{2}{*}{\textbf{Treatment}}   & \multicolumn{4}{c|}{\textbf{Mice and Elephant Flows}} & \multicolumn{4}{c|}{\textbf{1000 Genome Workflow}} & \multirow{2}{*}{\textbf{Median}} & \multirow{2}{*}{\textbf{IQR}} & \multirow{2}{*}{\textbf{\#BEST}} \\
\cline{3-10}
&  & \textbf{HAMILTON}	&	\textbf{BBR}	&	\textbf{RENO}	&	\textbf{CUBIC}	&	\textbf{HAMILTON}	&	\textbf{BBR}	&	\textbf{RENO}	&	\textbf{CUBIC} & & & 
\\\hline
\multirow{7}{*}{Recall} &
X-FLASH	&	\cellcolor{gray!20}84	&	\cellcolor{gray!20}50	&	\cellcolor{gray!20}89	&	\cellcolor{gray!20}70	&	\cellcolor{gray!20}85	&	\cellcolor{gray!20}92	&	\cellcolor{gray!20}73	&	\cellcolor{gray!20}95	&	85	&	17 & 8\\
& FLASH\_CART	&	81	&	\cellcolor{gray!20}49	&	84	&	\cellcolor{gray!20}68	&	71	&	82	&	66	&	87	&	76	&	15 & 2\\
& FLASH\_RF	&	80	&	\cellcolor{gray!20}49	&	84	&	\cellcolor{gray!20}68	&	78	&	85	&	52	&	91	&	79	&	20 & 2\\
& DODGE	&	78	&	48	&	85	&	\cellcolor{gray!20}68	&	70	&	81	&	63	&	86	&	74	&	15 & 1\\
& CART	&	81	&	\cellcolor{gray!20}49	&	84	&	\cellcolor{gray!20}68	&	72	&	82	&	67	&	87	&	77	&	14 &		2\\
& RF	&	80	&	\cellcolor{gray!20}49	&	84	&	\cellcolor{gray!20}68	&	77	&	85	&	52	&	91	&	79	&	20 &	2\\
& XGBOOST	&	71	&	39	&	74	&	58	&	69	&	81	&	39	&	89	&	70	&	22 &	0\\
\hline
\multirow{7}{*}{F-measure} &
X-FLASH	&	\cellcolor{gray!20}83	&	\cellcolor{gray!20}49	&	\cellcolor{gray!20}88	&	\cellcolor{gray!20}71	&	\cellcolor{gray!20}86	&	\cellcolor{gray!20}91	&	\cellcolor{gray!20}80	&	\cellcolor{gray!20}94	&	85	&	11 &	8\\
& FLASH\_CART	&	81	&	\cellcolor{gray!20}48	&	85	&	\cellcolor{gray!20}69	&	70	&	80	&	67	&	87	&	75	&	13 &	2\\
& FLASH\_RF	&	79	&	\cellcolor{gray!20}49	&	84	&	\cellcolor{gray!20}69	&	80	&	85	&	62	&	92	&	80	&	17 &	2\\
& DODGE	&	77	&	\cellcolor{gray!20}48	&	85	&	\cellcolor{gray!20}68	&	70	&	80	&	63	&	85	&	74	&	14 &	2\\
& CART	&	81	&	\cellcolor{gray!20}49	&	84	&	\cellcolor{gray!20}69	&	73	&	82	&	67	&	86	&	77	&	14 &	2\\
& RF	&	79	&	\cellcolor{gray!20}49	&	84	&	\cellcolor{gray!20}69	&	81	&	85	&	62	&	91	&	80	&	17 &	2\\
& XGBOOST	&	72	&	36	&	75	&	56	&	73	&	82	&	45	&	90	&	73	&	23 &	0\\
\hline
\multirow{7}{*}{G-score} &
X-FLASH	&	\cellcolor{gray!20}88	&	\cellcolor{gray!20}62	&	\cellcolor{gray!20}92	&	\cellcolor{gray!20}80	&	\cellcolor{gray!20}90	&	\cellcolor{gray!20}94	&	\cellcolor{gray!20}83	&	\cellcolor{gray!20}96	&	89	&	10 &	8\\
& FLASH\_CART	&	86	&	\cellcolor{gray!20}62	&	89	&	\cellcolor{gray!20}78	&	78	&	87	&	78	&	91	&	82	&	9 &	2\\
& FLASH\_RF	&	86	&	\cellcolor{gray!20}62	&	89	&	\cellcolor{gray!20}78	&	85	&	89	&	64	&	94	&	86	&	14 &	2\\
& DODGE	&	85	&	61	&	90	&	\cellcolor{gray!20}78	&	79	&	87	&	73	&	91	&	82	&	11 &	1\\
& CART	&	87	&	\cellcolor{gray!20}62	&	90	&	\cellcolor{gray!20}78	&	81	&	87	&	77	&	91	&	84	&	10 &	2\\
& RF	&	86	&	\cellcolor{gray!20}62	&	89	&	\cellcolor{gray!20}78	&	85	&	90	&	65	&	94	&	86	&	14 &	2 \\
& XGBOOST	&	80	&	52	&	83	&	70	&	79	&	88	&	52	&	93	&	80	&	18 &	0
\\

\hline
\end{tabular}
\vspace{-10pt}
\label{tbl:rq1}
\end{table*}

\noindent {\bf RQ1: Does tuning improve the performance of anomalies detection?} 

For our first set of results,
  default learners (RF and CART and XGBOOST) are compared with data miners with optimization (FLASH and DODGE ($\epsilon$)).

Table \ref{tbl:rq1} shows those results,
including the statistical ranking generated from  Scott-Knott test in \S IV.D for recall, F-measure, and G-score metrics defined in \S IV.C. Across all 8 datasets (HAMILTON, BBR, RENO, and CUBIC for 1000 Genome Workflow and Mice\&Elephant Flows), X-FLASH performed the best. Improvement with the previous work can be observed closely in Figure \ref{fig:delta}. X-FLASH improved up to 28\%, 29\%, and 40\% relatively for F-measure, G-score, and recall respectively. The benefit of tuning is even higher when comparing between XGBOOST and X-FLASH. In one extreme case, it improved 45\%, 52\%, \& 39\% to 80\%, 83,\%, \& 74\% respectively for F-measure, G-score, and recall (which are 60\%, 78\%, and 87\% relative improvement). This is explainable as shown in Table~\ref{tbl:params}. Except the number of estimators parameter (n\_estimators), the other three parameters values, when tuned, are far away from the default values. This shows that the default configurations for a data miner are not one-size-fits-all across different datasets and domains, hence, should be deprecated. With a mission-critical task like anomalies detection, it is essential to optimize the solution at hand specific for the domain, dataset, and metric. 

Moreover, X-FLASH notably outperformed DODGE($\epsilon$), where DODGE($\epsilon$) is a state-of-the-art data mining with optimization method taken from Software Engineering literature~\citep{dodge} (bug reports classification, close-issues prediction, defect prediction, etc). This shows that a method that works well for one disciplinary field may not work well in a different field. It is critical that the scientists and researchers revise and tune the method based on the specific conditions.

\begin{RQ}{\bf In term of correctness...}

X-FLASH was best at detecting anomalies. 
\end{RQ}

\begin{table}[!b]
\footnotesize
\begin{center}
\caption{Sorted time performance (in seconds) on median for  the studied data mining methods on eight datasets, across TCP variants (H-Hamilton, B-BBR, R-Reno, and C-Cubic)  for  Mice\&Elephant Flows and 1000 Genome Workflow.}
\vspace{5pt}
\label{tbl:time}
\hspace{-10pt}\begin{tabular} {p{.14\linewidth}|c|c|c|c|c|c|c|c}
&  \multicolumn{4}{ c| }{Mice\&Elephant Flows} & \multicolumn{4}{ c }{1000 Genome Workflow} \\ \hline
 Type & H & B & R & C & H & B & R & C \\ \hline
 XGBOOST  & 6 & 5 & 5 & 5 & 16 & 4 & 15 & 4    \\ \hline
 DODGE($\epsilon$) & 130 & 167 & 173 & 272  & 63 & 24 & 154 & 32   \\ \hline
 X-FLASH & 628 & 400 & 437 & 563 & 453 & 139 & 617 & 165  \\ 
\end{tabular}
\end{center}
\vspace{-15pt}
\end{table}

\noindent
{\bf RQ2: Is tuning anomalies detection impractically slow?} 

To our surprise, X-FLASH achieved statistically significant improvement in the performance scores for our data miners in less than 30 evaluations.  In this space, our proposed solution took the most time among default and the state-of-the-art optimizer DODGE($\epsilon$) (30 evaluations). However, considering the mission-critical nature of the problem and it still take less than 11 minutes at most with standard hardware (i.e., CPU) from Table~\ref{tbl:time}.  The performance increments seen in Figure~\ref{fig:delta} and Table~\ref{tbl:rq1} are more
than to compensate for the extra CPU required for X-FLASH. Modern hardware choices (e.g., GPUs) and parallel computation can be configured to improve the time to be more practical for the industry.

\begin{RQ}{\bf In term of runtime performance...}

X-FLASH was the worst but still converged in less than 11 minutes at max. Hence, the cost of running X-FLASH
worths the performance improvement.
\end{RQ}


\begin{table}[!b]
\begin{center}
\scriptsize
\caption{Non-overlapped Features selected by default XGBOOST versus after tuned by FLASH across TCP variants (H-Hamilton, B-BBR, R-Reno, and C-Cubic)  for  Mice\&Elephant Flows and 1000 Genome Workflow.}
\label{tbl:rq3}
\begin{tabular}{l|l|l|l}
\multicolumn{2}{ c| }{\textbf{Datasets}} & \textbf{DEFAULT} & \textbf{Tuned by FLASH} \\ \hline
Mice & H & s\_rtt\_avg, s\_ack\_cnt\_p   &  c\_ttl\_min, s\_rtt\_min \\ 

 \& & B & c\_bytes\_uniq, s\_rtt\_avg & c\_fin\_cnt, c\_first\_ack \\
Elephant & R & s\_ack\_cnt\_p  & c\_ttl\_min \\
Flows &  C & s\_rtt\_avg, s\_ack\_cnt\_p & c\_first\_ack, s\_rtt\_max \\\hline
& H & s\_last\_handshakeT, c\_rtt\_std,   &  s\_pkts\_retx, s\_pkts\_data, \\ 
   &  & c\_pkts\_unfs, c\_ack\_cnt\_p & s\_fin\_cnt, s\_rtt\_cnt \\ 

 & B & c\_appdataT, s\_first\_ack,  & c\_pkts\_reor, c\_bytes\_retx,  \\
1000  & & s\_win\_max, s\_rtt\_std & c\_win\_max, c\_pkts\_fs \\

Genome & R & c\_bytes\_retx, c\_first\_ack  & c\_pkts\_rto, c\_pkts\_retx,  \\
Work- & &  s\_first\_ack, c\_mss\_max & c\_pkts\_ooo, s\_win\_min,  \\
Flow & &  s\_win\_max, c\_appdataT & c\_pkts\_unk, c\_cwin\_max \\

& C & c\_pkts\_rto, c\_cwin\_max,  & c\_pkts\_fs, s\_cwin\_min, \\
&  & s\_rtt\_min, c\_appdataT,  & c\_pkts\_unfs, c\_bytes\_retx \\
& & s\_ack\_cnt\_p & c\_ack\_cnt\_p \\

\end{tabular}
\vspace{-15pt}
\end{center}
\end{table}

\noindent
{\bf RQ3: Does tuning change conclusions about what factors are most important in anomalies detection?}  

It is important to understand which attribute(s) associated more with differentiating characteristics between different types of anomalies and normal scientific flows. Scientists and network managers can then inspect the flagged ones with high likelihood of anomalies. From Table \ref{tbl:rq3},  among the top ten important features (curated from the built-in feature\_importances\_ function \cite{scikit-learn}), the median number of features is seven as commonly chosen decisive factors for the anomalies detectors (while the rest 30\% of top ten features are not the same, non-overlapped features). It demonstrated how the previous study~\cite{tcp_indis_19} and conclusion from default learner can be untrustworthy.  They did not attempt to do the features importance analysis of their anomaly detectors. 

Interestingly, the learners rankings were also changed slightly with tuning. Default CART model performed similarly or better than default XGBoost in 7 out of 8 datasets across recall, F-measure, and G-score respectively. However, after tuning through FLASH, XGBoost always better across 8 datasets for each metric.   

\begin{RQ}{\bf In term of real-world applicability...}

Tuning has shown how the features are considered important differently with and without it which can negatively affect the real-world.
\end{RQ}


\section{Discussions}

In this section, we discuss the possible factors that  affect the effectiveness of evaluations. 
Such factors also commonly exist in other research works with large scale empirical studies.


\subsection{Evaluation Bias}

This paper employed recall, F-measure, and G-score to evaluate the overall performance. We have taken into generalization issues of single metrics (e.g., accuracy and precision) into consideration and instead evaluate our methods on metrics that aggregate multiple metrics like F-measure and G-score.  As the future work, we plan to test the proposed methods with additional analysis that are endorsed within SE literature (e.g., P-opt20~\cite{Tu20_emblem}) or general ML literature (e.g., MCC~\cite{mcc_metrics}).

For result validity, we applied  bootstrap significant test and the cliff-delta effect size test. Hence, in this paper, ``X was different from Y'' conclusions were based on both tests.

\subsection{Learner Bias}

This work proposed X-FLASH (XGBoost + FLASH) data mining method and compared it with DODGE ($\epsilon$), endorsed by the SE literature. 
As the future work, we plan to test if the conclusions (data miner + optimization, called X-FLASH, is a good way to detect and classify anomalies) hold across multiple tasks associated with scientific workflows.

\subsection{Sampling Bias}

As a common issue for data mining field,
our work is subject to possible sampling bias, i.e., the conclusion for the data we studied in this paper may
not hold for other types of data. 
To ensure data and code availability for the research community, we release our code and data at \url{https://github.com/msr2021/tuningworkflow/}. 


\subsection{External Validity}

The approach, DODGE($\epsilon$), that SE literature have established as 
``standard tool'' may not be ``general'' to all the fields. Rather, the tools that are powerful in their home domain may need to be used with caution, if applied to new domains such as scientific workflow (and specifically, anomalies detection). Some of the data quirks that essential to the success of DODGE($\epsilon$) include: (1) the prediction is binary (e.g., 0 or 1, faulty or non-faulty, etc); and (2) the target class is infrequent. Those data quirks may lead to issues such as:

\bi
\item It is harder to find the target;
\item The larger the  observed  $\epsilon$ in the results;
\item The greater the number of redundant tunings;
\ei 

Therefore, since software engineering often deals
with relatively infrequent target classes, we should expect to see a large $\epsilon$ uncertainty in our conclusion which is more likely that DODGE($\epsilon$) will work. However, for the faulty TCP file transfers detection in this paper, it is a multiclass classification and the distribution is only infrequent for the normal flows instead of the targets (i.e., anomalies flows). Therefore, for such a more dynamic problem with multiple target classes and the distribution is diverse, FLASH is recommended as a more general approach for optimization. 

In summary, and in support of the general theme of this paper, this external validity demonstrates the danger of treating
all data with the state-of-the-art method, especially when switching domain (e.g., DODGE($\epsilon$) from SE literature). 

\section{Conclusion}

In this paper, we show that utilizing general anomalies learning tools for faulty TCP file transfers without tuning can be considered \textit{harmful} and \textit{misleading} to the reliability of networked infrastructures. 
Our proposed solution X-FLASH combined an ensemble model (XGBoost) and a sequential model-based optimizer (FLASH) from Software Engineering literature to detect and classify the correct malicious activity or attacks, before it
contaminates downstream scientific process:

\bi 
\item Tuning default learners will improve the relative performance up to 28\%, 29\%, and 40\% for F-measure, G-score, and recall (see Table \ref{tbl:rq1}) from the SOTA work~\cite{tcp_indis_19}. 
\item Tuning changes previous conclusions on what learner is the best performing, i.e., from RF to XGBoost.

\item Tuning changes previous conclusions on what  factors are most
influential in detecting for anomalies by 30\% (see Table \ref{tbl:rq3}). 
\ei

Moreover, results showed that X-FLASH out-performed state-of-the-art data mining in SE literature, DODGE($\epsilon$) by ~\citet{dodge}. This result is suggestive (but not conclusive) evidence that (a)~prior work on analytics has {\em  over-fitted} methods (to systems like Apache);  and that (b)~there is {\em no better time} than now to develop new case studies (like scientific workflows).

As to future work, it is now important to explore the implications of these conclusions to other kinds of scientific workflow analytics. Specifically, previous papers for anomalies detection for scientific workflow~\citep{mandal16, gaikwad16, temporal18, herath19, samak11, herath19, buneci08,BALA2015980, samak13} that not based on TCP data transfers should be also reinvestigated as none have done tuning study to avoid falling in the same \textit{retracted} category.

\section{Acknowledgements}


We thank the Computational Science community
from the Pegasus Research group and Renaissance Computing Institute at UNC (RENCI) for
their assistance with this work. 

This work was partially funded by 
an NSF CISE Grant \#1826574, \#1931425 and DOE contract number \#DE- SC0012636M, ``Panorama 360: Performance Data Capture and Analysis for End-to-end Scientific Workflows''.

\balance
\bibliographystyle{IEEEtranN}

\bibliography{cas-refs}





\end{document}